\documentclass[twocolumn,showpacs,preprintnumbers,amsmath,amssymb]{revtex4}
\usepackage[T1]{fontenc}  
\usepackage[english]{babel}
\usepackage[dvips]{graphics}

\newcommand{\beq}{\begin{equation}}
\newcommand{\eeq}{\end{equation}}
\newcommand{\bma}{\begin{math}}
\newcommand{\ema}{\end{math}}
\newcommand{\beqa}{\begin{eqnarray}}
\newcommand{\eeqa}{\end{eqnarray}}

\def\opone{\le\textbf{}\textbf{}avevmode\hbox{\small1\kern-3.8pt\normalsize1}}

\begin{document}

\title{One-Dimensional Theory of the Quantum Hall System}

\author{Emil J. Bergholtz}

\email{ejb@physto.se}

\author{Anders Karlhede}

\email{ak@physto.se}

\affiliation{Department of Physics,
Stockholm University \\ AlbaNova University Center\\ SE-106 91 Stockholm,
Sweden}

\date{\today}

\begin{abstract} 
We consider the lowest Landau level on a torus as a function of its circumference $L_1$.  
When $L_1\rightarrow 0$, the ground state at general rational filling fraction is a crystal with a gap---a 
Tao-Thouless state. For filling fractions $\nu=p/(2pm+1)$, these states are the limits of Laughlin's or Jain's wave functions 
describing the gapped quantum Hall states when $L_1\rightarrow \infty$. For the half-filled Landau level, there is a transition to 
a Fermi sea of non-interacting neutral dipoles, or rather to a Luttinger liquid modification thereof, at $L_1\sim5$ magnetic lengths. This state is a 
version of the Rezayi-Read state, and develops continuously into the state that is believed to describe the observed metallic phase 
as $L_1\rightarrow \infty$.  Furthermore, the effective Landau 
level structure that emerges within the lowest Landau level follows from the magnetic symmetries.

\end{abstract}


\maketitle

Laughlin's \cite{Laughlin83} and Jain's \cite{Jain} wave functions for the fractional quantum Hall states and 
Jain's radical interpretation 
of these in terms of non-interacting particles (''composite fermions'') \cite{Jain98}, whereby the fractional quantum Hall effect 
becomes an integer effect of composite fermions 
in a reduced magnetic field have been extremely successful. Mean field theories of the Chern-Simons type, where the external 
magnetic flux is (partially) cancelled by a smeared statistical flux, have been developed \cite{cs}. 
The half-filled Landau level is supposedly
mapped onto free particles in no magnetic field. The mean field theory of this state, due to Halperin, Lee and 
Read \cite{hlr}, predicted a Fermi momentum and a momentum dependence in the conductivity that was  
confirmed by the surface acoustic wave experiments by Willett {\it et al} \cite{Willett90}. 
Furthermore, ballistic transport showed particles moving in agreement with the composite 
fermion prediction \cite{ballistic}.
The theory of the half-filled Landau level was further developed by several groups and a description in terms of neutral dipoles was 
proposed \cite{several}.  Exact diagonalization of small systems not only supports the correctness of 
Laughlin's and Jain's wave functions, but also indicate an emerging Landau level structure within the 
lowest Landau level, in the sense that the low energy states of the interacting 
electron system resemble the states for free fermions in a reduced magnetic field. 

In spite of the spectacular success of the description of the quantum Hall system in terms of composite fermions,
a  sound theoretical foundation for them is,  in our opinion, still lacking \cite{Dyakonov}.  Although they were 
motivated by the form of Laughlin's and Jain's wave functions it has not been possible to make this connection precise. 

We recently obtained an exact solution describing the interacting electron gas in the half-fillled lowest 
Landau level on a thin torus \cite{Bergholtz05}:  
The ground state, which is homogeneous, is a filled Fermi sea of non-interacting neutral dipoles and the excitations are gapless. 
In this Letter we identify this ground state with the Rezayi-Read state \cite{rr} and conclude that the exact 
solution develops continuously, without a phase transition, into the two-dimensional bulk state.   

For spin-polarized electrons at  a generic filling fraction $\nu=p/q\le1$, we find that the ground state is a regular lattice of 
electrons determined by electrostatics alone when $L_1\rightarrow 0$, $L_1$ being 
the (short) circumference of the torus. This crystal has a gap---the lattice is rigid and there are no phonons. 
For $\nu=1/q$ this state is the Tao-Thouless (TT) state \cite{Tao83} 
with one electron on every $q$:th site, and for odd $q$ it develops continuously, remaining the ground state for a short-range interaction,
into the Laughlin state as $L_1\rightarrow \infty$ \cite{Haldane94}. We call these gapped crystals TT states for general $\nu=p/q$. 
The TT state has, in spite of being spatially inhomogeneous, the same (magnetic) symmetries as the homogeneous Laughlin state, it also 
has excitations with the same fractional charge. 
At $\nu=p/(2pm+1)$ we find that the TT states are the $L_1\rightarrow 0$ limits of the Jain states, and 
the fractionally charged excitations are domain walls separating degenerate ground states. 
We suggest that the gapped quantum Hall states in general develop continuously 
from the TT ground states to the
bulk states as $L_1\rightarrow \infty$. However, for $\nu=1/2$, we find a phase transition at $L_1\approx 5.3$ magnetic lengths to the  
gapless system of neutral dipoles, or rather to a Luttinger liquid modification thereof, which then develops continuously 
into the bulk state as $L_1\rightarrow \infty$. Transitions from the small $L_1$ TT state to other states may occur as $L_1$ increases. 
For example, at $\nu=1/q$, $q$ odd and large one expects a transition to a gapless Wigner crystal \cite{Lam}. We conclude that $L_1$ is a parameter 
of a controlled and systematic expansion  around a solvable case---the thin torus, or equivalently, a short range spin chain. 

We show that the degeneracy of the quasiparticle states at $\nu=p/(2pm+1)$, and the effective Landau 
level structure that is observed to emerge in the lowest Landau level, is a consequence of the $q$-fold degeneracy of 
the states at filling factor $\nu=p/q$ \cite{Haldane85PRL}, {\it ie} it is a consequence of the magnetic translation symmetries. 
We relate this to the composite fermion interpretation. 

We introduce the one-dimensional model by briefly discussing the lowest Landau level on a torus with lengths $L_1,L_2$ in the $x$ and 
$y$-directions respectively \cite{Haldane85, Haldane85PRL}; we use units where $\hbar= c/eB=1$.  In 
Landau gauge, ${\bf A}=By\hat {\bf x}$, the magnetic translation operators $t_{\alpha}, \alpha=1,2$ that translate an electron a distance 
$L_{\alpha}/N_s$ in the $\alpha$-direction are $t_1=e^{(L_1/N_s)\partial _x}, \, t_2=e^{(L_2/N_s)(\partial _y +ix)}$, 
where $N_s=L_1 L_2/(2 \pi)$ is the number of flux quanta through the surface. 
The states $\psi_k=t_2^k \psi_0$, $k=0,1,...N_s-1$,  
where $\psi_0=\pi^{-1/4}L_1^{-1/2}\sum_n e^{inL_2x} e^{-(y+nL_2)^2/2}$, is a basis in the lowest Landau level. $\psi_k$ is located along the line 
$y=-2\pi k/L_1$ and is a  $t_1$ eigenstate, $t_1\psi_k=e^{i2\pi k/N_s}\psi_k$. Letting $c^{\dagger}_k$ create an electron in state $\psi_k$, 
$\{c_k,c_m^\dagger \}=\delta_{km}$, maps the lowest Landau level onto a one-dimensional lattice model with lattice constant 
$2\pi/ L_1$. An $N$-particle state is characterized by the positions $\{k_1,k_2,...k_N\}$ of the particles.
The general translationally invariant two-body Hamiltonian becomes $H=\sum_n \sum_{k>m} V_{km}c^\dagger_{n+m}c^\dagger_{n+k}c_{n+k+m}c_n$, where
$V_{km}=V_{k,-m}\ge0$. $V_{km}$ is the amplitude for two electrons separated $k-m$ lattice constants to hop symmetrically to a separation 
$k+m$ lattice constants. In particular, $V_{k0}$ describe the electrostatic repulsion (including the exchange interaction). 
For a given real space interaction $V(r)$, the interaction is dominated by the terms with small $k,m$ as the torus becomes thin 
($L_1\rightarrow 0$) since the lattice constant $2\pi/L_1\rightarrow \infty$. In the two-dimensional limit, $L_1\rightarrow \infty$, the 
range of the interaction measured in terms of lattice constants goes to infinity.

Consider the electron gas at filling fraction $\nu=p/q$, where $p$ and $q$ are relatively prime integers and the number of 
electrons, $N_e=N_s p/q$, is an integer. The operators $T_{\alpha}=\prod_i^{N_e} t_{i\alpha}$, ($t_{i\alpha}$ translates electron $i$) 
commute with $H$ and, since $T_{\alpha}^{N_s}=1$, the eigenvalues are 
$e^{2\pi iK_{\alpha} /N_s}, K_{\alpha}=0,...N_s-1$. $\{H, T_1, T_2^q\}$ is a maximally commuting set of operators. $T_2$ changes $K_1$ by $N_e$
and leaves the energy unchanged. Hence, each energy eigenstate is $q$-fold degenerate and we choose to characterize it  by the smallest $K_1$.
Thus, the energy eigenstates are characterized by a two-dimensional vector $K_\alpha=0,...,N_s/q-1$ \cite{Haldane85PRL} \footnote{$K_\alpha$ is 
in one to one correspondence with  $k_\alpha$ in Ref. \cite{Haldane85PRL}; $k_\alpha$ characterizes the relative 
motion of the electrons.}. 

At $\nu=1/2$, we obtained an exact mapping of the low energy sector onto free neutral fermions (dipoles) for  the short range interaction 
$V_{km}=V^*_{km}$ with non-vanishing elements $V^*_{21}, \,V^*_{10}=2V^*_{20} $ \cite{Bergholtz05}. 
The ground state is a homogeneous one-dimensional Fermi sea of these 
neutral dipoles and the excitations are gapless particle-hole excitations out of this sea (when $N_e\rightarrow \infty$).  $V^*_{km}$ is a good approximation 
to a real space short-range repulsion for $L_1\sim 5$.  A renormalization 
group argument showed that the ground state develops into an interacting Luttinger liquid near $V^*_{km}$. Based on the similarity of the 
obtained state to what is expected for the bulk system---a homogeneous gapless ground state made out of free neutral 
dipoles---we conjectured that this state develops continuously, without a phase transition, to the bulk two-dimensional state as 
$L_1 \rightarrow \infty$.

To investigate this conjecture we have performed exact diagonalization of up to $N_e=10$ electrons for various  $L_1$ using an unscreened Coulomb interaction. 
A typical phase diagram is 
shown in Fig. 1. Independent of $N_e$, we find that the ground state is the TT state 01010101.... for  $0\le L_1 \lesssim 5.3$. 
At $L_1\approx 5.3 $ there is a sharp transition to a state that we identify as the ground state of the solvable model $V^*_{km}$ by comparing the quantum 
numbers and calculating the overlap. 
When $L_1$ increases further, a series of transitions to new states occurs. The number and position of these transitions, 
which are very smooth compared to the one at $L_1\approx 5.3$, depend on $N_e$.

\begin{figure}[h!]
\begin{center}
\resizebox{!}{23mm}{\includegraphics{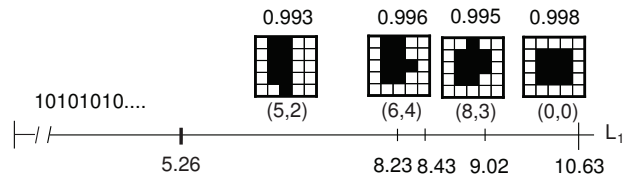}}
\end{center}
\caption{\textit{{\small Phase diagram for $\nu=1/2$ obtained in exact diagonalization for 9 electrons with
Coulomb interaction. For $L_1\lesssim 5.3$ the state is 101010... as 
indicated, whereas for $L_1\gtrsim 5.3$ the number gives the overlap with the Rezayi-Read state with the displayed Fermi sea of momenta ${\bf k}_i$; the 
variation of the overlap is small within each region.  
The quantum numbers $(K_1,K_2)$ are shown below each Fermi sea. (The 
phase diagram is symmetric about  $L_1=L_2=\sqrt{4\pi N_e}$).
}
}}
\end{figure}

The bulk $\nu=1/2$ state is believed to be described by the Rezayi-Read wave function \cite{rr}, which on the torus 
takes the form \cite{rrhaldane} 
\begin{eqnarray}
\Psi_{RR}  = {\rm det_{ij}}[e^{i{\bf k}_i\cdot {\bf R}_j}] \Psi_{\frac 1 2}  \ \ \ ,
\end{eqnarray}
where $R_{ix}=(2\pi/L_1)(x_i-i\partial_{y_i}), \, R_{iy}=(2\pi /L_2)i\partial_{x_i}$, 
and $\Psi_{\frac 1 2}$ is the bosonic Laughlin state at $\nu=1/2$.
This wave function depends on a set of ''momenta'' $\{{\bf  k}_i \}$, which determine the 
conserved quantum numbers $K_\alpha=\sum_i k_{i\alpha}$. For each  
ground state we obtained by exact diagonalization for $L_1\gtrsim 5.3$ and odd $N_e\le 9$ we found a choice of $\{{\bf k}_i \}$ such that the overlap between 
the ground state and this Rezayi-Read state is close to one \footnote{For even $N_e$, the ground state is a linear 
combination of two Rezayi-Read states with diffferent sets of  $\{{\bf k}_i \}$.}. 
Moreover, the ''Fermi sea'' of momenta $\{{\bf  k}_i \}$ develops in a systematic way from 
an elongated sea to a ''circular'' one as $L_1$ increases, as is shown in Fig. 1 for $N_e=9$. 

Our interpretation of this is that the state obtained for $V^*_{km}$ does indeed develop continuously into the bulk state as $L_1 \rightarrow \infty$ 
without any phase transition. The level crossings observed for $L_1 > 5.3$ are to similar states. Note that the bulk
two-dimensional state is expected to be gapless, thus level crossings may occur when the thermodynamic limit is taken. 
An interesting question concerns the dimensionality of the system. 
For $L_1$ slightly larger than 5.3 and $L_2\rightarrow \infty$
the system is a Luttinger liquid, {\it ie} a one-dimensional gapless state. This is described by the Rezayi-Read wave function with momenta 
forming an elongated Fermi sea. When $L_1$ 
increases these momenta change gradually---only one ${\bf  k}_i$ changes at each level crossing as in Fig. 1. This strongly suggests that  the 
state remains gapless (for $L_1$ finite, $L_2\rightarrow \infty$), in which case it should still be a Luttinger liquid. The question is what happens 
when $L_1\rightarrow \infty$. Does the state remain a Luttinger liquid or does it become a free two-dimensional Fermi gas as the standard 
composite fermion theory predicts? 

We now consider general rational filling fractions in the lowest Landau level. When $L_1 \rightarrow 0$, the hopping terms in the 
interaction vanish exponentially and only the electrostatic terms 
$V_{k0}$ remain \cite{Westerberg}. The eigenstates are then states with fixed charges.
For  a reasonable repulsive interaction such as the Coulomb interaction, the ground state is a crystal, the TT state,  
that separates the charges as much as possible and there is a gap to excitations. 
For example, for the Jain fraction  $\nu=p/(2pm+1)$, the unit cell of the TT state is $10_{2m}(10_{2m-1})_{p-1}$, in obvious chemical notation, 
whereas for $\nu=1/2m$ it is $10_{2m-1}$.  For $\nu=1/(2m+1)$, the TT state is the limit of the Laughlin state 
as mentioned above, whereas for $\nu=1/2$ it is the  ground state for $L_1\lesssim 5.3$ \cite{chui86}, see Fig. 1. 

Generalizing Rezayi and Haldane \cite{Haldane94}, we consider the $L_1 \rightarrow 0$ limit of the Jain state at filling fraction $\nu=p/(2pm+1)$.
On the cylinder it  takes the form $\Psi_{\frac p {2pm+1}}={\cal P}_{LLL} \Phi_p J^{2m} e^{-\sum y^2/2}$,
where ${\cal P}_{LLL}$ denotes projection onto the lowest Landau level, $\Phi_p$ is $p$ filled Landau levels and $J$ is the Jastrow factor
$J =\prod_{i<j}(e^{2\pi i z_i/L_1}-e^{2\pi i z_j/L_1}) $ \footnote{In the following,  we do not impose periodicity in the $y$-direction, 
this makes no difference when $L_2\rightarrow \infty$.}. 
Expanding $\Psi_{\frac p {2pm+1}}$ as a 
power series in $e^{2\pi i z_i/L_1}$ and expressing the result in terms of the single-particle wave functions
$\psi_k=\pi^{-1/4}L_1^{-1/2}e^{2\pi i k z/L_1}e^{-y^2/2}e^{-2\pi ^2 k^2/L_1^2}$ one reads off the states $\{k_1,k_2,...k_N\}$ in the 
particle number basis on the one-dimensional lattice and finds that they are 
multiplied by the factor $e^{\sum_i (2\pi k_i/L_1)^2/2}$. $\Psi_{\frac p {2pm+1}}$ consists of a set of states, 
which all have the same $K_1=\sum_ik_i$, 
and when $L_1  \rightarrow 0$ the component with largest  $\sum_i k_i^2$ dominates. 
We find that $\Psi_{\frac p {2pm+1}}$ approaches the TT state with unit cell 
$10_{2m}(10_{2m-1})_{p-1}$.
For example, the sequence of states that approaches $\nu=1/2$ from below have unit cells: 100,10010,1001010,100101010,....  
These limits of the Jain states are identical to the states determined by the electrostatic repulsion on the thin torus above. 
Thus we conclude that the Jain states are the ground states on a thin torus.  It is only for the Laughlin states with a short-range 
interaction that it is established that no phase transition occurs as a function of $L_1$ \cite{Haldane94}. However, it seems likely that this is 
true for Jain states as well.  This is supported by exact diagonalization: We have considered all $\nu=p/q\le 1$ with $q\le 11$ and find, for
Coulomb interaction, a transition if and only if $q$ is even. 

At $\nu= p/(2p+1)$, electrostatics and hopping cooperate---the TT ground state for small $L_1$ minimizes the electrostatic repulsion and is 
a very  ''hoppable'' state in the sense that $V_{21}$ generates many states when acting repeatedly on it. There is no phase transition as $L_1$ grows.
At $\nu=1/2$, on the other hand, the TT ground state at small $L_1$ is annihilated by $V_{21}$. Electrostatics and hopping compete, 
leading to a phase transition as $L_1$ grows. 

The fractional quantum Hall state has  excitations with fractional charge. In the $L_1 \rightarrow 0$ limit, the quasihole (quasielectron) 
at $\nu=p/(2pm+1)$ is created by removing  (inserting) $10_{2m-1}$ somewhere in the TT state with unit cell 
$10_{2m}(10_{2m-1})_{p-1}$ \footnote{For $p=1$, a quasihole is created by inserting a zero since 
the unit cell is $10_{2m}=10_{2m-1}0$---an  operation that is closely related to Laughlin's original creation of a quasihole by 
inserting a flux quantum \cite{Laughlin83}. }.
These quasiparticle states, which can be interpreted as domain walls separating degenerate ground states,
are the $L_1 \rightarrow 0$ limits of those obtained by moving a composite fermion from a filled to an empty Landau level \cite{Jain98}. 
The charge of the excitations can be read off directly in the $L_1\rightarrow 0$ limit using Schrieffer's counting
argument \cite{Schrieffer}: Removing $2pm+1$ widely separated $10_{2m-1}$ and adding $2m$ unit cells to keep the 
number of sites unchanged creates $2pm+1$ quasiholes, each with charge $e^*=e((2pm+1)-2mp)/(2pm+1)=e/(2pm+1)$. 

We conclude that the TT states at small $L_1$ are limits of quantum Hall states for all the Jain fractions and 
suggest that this is the case also for more general gapped quantum Hall states. An example of such are the recently  observed minima 
in $R_{xx}$ at fractions $\nu=5/13,3/8,4/11,6/17$ in the interval $[1/3,2/5]$ \cite{Pan03}. 
The $L_1\rightarrow 0$ ground state unit cells at these fractions are $(10_2)_{2}1010_2 10$ and $10(10_2)_{p-1}$, $p=3,4,6$. 
For 3/8 we find transitions as $L_1$ increases, whereas no transition is found for 4/11, 
indicating that the former is gapless while the latter is gapped.

We now show that an effective Landau level structure emerges within the lowest Landau level on the torus as a consequence of symmetry. 
Consider filling factor $\nu=p/(2pm+1)$ with $N_e=pN$ electrons. $N$ is the number of unit cells in the TT ground 
state when $L_1 \rightarrow 0$. Imagine creating 
a quasielectron by inserting $10_{2m-1}$ somewhere in the TT state. There are $N$ degenerate orthogonal quasielectron 
states, {\it ie} one per unit cell \footnote{Here we assume 
that at least one electron is kept fixed relative to the ground state, this divides out the $2pm+1$-fold ground state degeneracy.}. 
At the new filling factor $\nu =(pN+1)/((2pm+1)N+2m)$, the center of mass degeneracy of any state is 
$(2pm+1)N+2m$ \cite{Haldane85PRL}; subtracting the number of added sites $2m$
and dividing by the ground state degeneracy $2pm+1$, gives the number of degenerate quasielectron states $N$. Thus the 
degeneracy of the quasielectron state is a consequence of symmetry and hence holds for any $L_1$---provided the  quasielectron 
is created by adding $2m$ sites and one electron. 

The $N=N_e/p$ degenerate
quasielectron states at $\nu=p/(2pm+1)$, form, in composite fermion language, the $p+1$:st composite fermion Landau level. Filling this, 
by adding $N$ quasielectrons, gives $\nu=(pN+N)/((2pm+1)N+2mN)=(p+1)/(2(p+1)m+1)$, {\it ie } the next state in the Jain hierarchy. Hence this
state can be interpreted as consisting of  $p+1$ filled composite fermion Landau levels---if the $p=1$ state is interpreted as a filled 
Landau level of $10_{2m}$, {\it ie} of electrons bound to $2m$ holes. Note however, that the process of successively 
creating quasielectrons by inserting $10_{2m-1}$ also makes explicit, on the thin torus, the original hierarchy construction due to Haldane and 
Halperin \cite{hierarchy}. Starting from the Laughlin state at $\nu=1/(2m+1)$ with unit cell $10_{2m}$,  
this eventually creates the $\nu=2/(4m+1)$ state with unit cell  $10_{2m}10_{2m-1}$. Continuing the process of 
inserting $10_{2m-1}$ one successively obtains all the states $\nu=p/(2pm+1)$.  Whereas the operation of inserting  $10_{2m-1}$ is unique and 
well-defined through this process, its interpretation depends on $p$: near $\nu=p/(2pm+1)$ it corresponds to creating a quasielectron 
with charge $-e/(2pm+1)$.

We have investigated a one-dimensional theory for the lowest Landau level and derived an emergent
Landau level structure from symmetry. The theory has a  
dimensionless parameter $L_1$, and the two-dimensional bulk quantum Hall system is obtained when  $L_1\rightarrow \infty$. 
A one-dimensional formulation is natural, and in a sense obvious, since a single Landau level is 
a one-dimensional system. What is non-trivial is that a one-dimensional lattice model with short-range interaction is useful. 
As $L_1\rightarrow 0$, 
the ground state at filling factor $\nu=p/q$ is a gapped crystal, the TT state, determined by electrostatics alone.  We have argued that for the Jain filling factors, 
and presumably also for more general odd $q$ states, these TT states develop continuously to the quantum Hall states 
when $L_1\rightarrow \infty$. For $\nu=1/2$, we found a phase transition to a Luttinger liquid of neutral dipoles (composite fermions) 
for small $L_1$ and showed that this state develops continuously to the metallic state as $L_1\rightarrow \infty$.  At other 
filling fractions, and also depending on the interaction, transitions to other states may occur.

We thank Thors Hans Hansson for numerous valuable discussions. This work was supported by the Swedish Research Council and by NordForsk.


\begin{thebibliography}{99}

{\footnotesize
\bibitem{Laughlin83} R.B. Laughlin,  Phys. Rev. Lett. {\bf 50}, 1395 (1983); R.B. Laughlin in {\it The Quantum Hall Effect}, 
eds. R.E. Prange, and S.M. Girvin, (Springer-Verlag, New York, 1990).

\bibitem{Jain} J.K. Jain,  Phys. Rev. Lett. {\bf 63}, 199 (1989).

\bibitem{Jain98} For reviews, see, J.K. Jain in {\it Perspectives in Quantum Hall Effects}, eds. S. Das Sarma, and A. Pinczuk, (John Wiley \& Sons, New York, 1996);  
J.K. Jain, and R.K. Kamilla, in {\it Composite Fermions}, 
ed. O. Heinonen, (World Scientific, Singapore, 1998) and other articles therein.

\bibitem{cs} S-C. Zhang, T.H. Hansson, and S.A. Kivelson, Phys. Rev. Lett. {\bf 62}, 82 (1988); N. Read, Phys. Rev. Lett. {\bf 62}, 86 (1988);
A. Lopez, and E. Fradkin, Phys. Rev. B {\bf 44}, 5246 (1991); {\it ibid.} {\bf 47}, 7080 (1993). 

\bibitem{hlr} B.I. Halperin, P.A. Lee, and N. Read, Phys. Rev. B {\bf 47}, 7312 (1993); see also V. Kalmeyer, and S.-C. Zhang,  
Phys. Rev. B {\bf 46}, R9889 (1992).

\bibitem{Willett90} R.L. Willett, M.A. Paalanen, R.R. Ruel, K.W. West, L.N. Pfeiffer, and D.J. Bishop,   Phys. Rev. Lett. {\bf 65}, 112 (1990).

\bibitem{ballistic} W. Kang, H.L. St\"ormer, L.N. Pfeiffer, K.W. Baldwin,  and K.W. West, Phys. Rev. Lett. {\bf 71}, 3850 (1993); V.J. Goldman, B. Su, and J.K. Jain, 
Phys. Rev. Lett. {\bf 72}, 2065 (1994); J.H. Smet, D. Weiss, R.H. Blick, G. L\"{u}tjering, K. von Klitzing, R. Fleischmann, R. Ketzmerick, T. Geisel, and G. Weimann, 
Phys. Rev. Lett. {\bf 77}, 2272  (1996).

\bibitem{several} G. Murthy, and R. Shankar, in {\it Composite Fermions}, ed. O. Heinonen, (World Scientific, Singapore, 1998) 
and Rev. Mod. Phys. {\bf 75}, 1101 (2003);
D.-H. Lee, Phys. Rev. Lett. {\bf 80}, 4745 (1998), {\bf 82}, 2416(E) (1999); N. Read, Phys. Rev. B {\bf 58}, 16262 (1998);
V. Pasquier, and F.D.M. Haldane,  Nuclear Physics B {\bf 516}, 719 (1998).

\bibitem{Dyakonov} For a critical review, see M.I. Dyakonov, in
{\it Recent Trends in Theory of Physical Phenomena in High Magnetic Fields}, eds. I.D. Vagner, P. Wyder, and T. Maniv, (Kluwer, 2003). 


\bibitem{Bergholtz05} E.J. Bergholtz, and A. Karlhede, Phys. Rev. Lett. {\bf 94}, 26802 (2005).

\bibitem{rr} E.H.  Rezayi, and N. Read,  Phys. Rev. Lett. {\bf 72}, 900  (1994).

\bibitem{Tao83} R. Tao, and D.J. Thouless, Phys. Rev. B {\bf 28}, 1142 (1983).

\bibitem{Haldane94}  E.H. Rezayi, and F.D.M. Haldane, Phys. Rev. B {\bf 50}, 17199 (1994).

\bibitem{Lam} P.K. Lam, and S.M. Girvin, Phys. Rev. B {\bf 30}, 473 (1984).

\bibitem{Haldane85PRL} F.D.M. Haldane, Phys. Rev. Lett. {\bf 55}, 2095 (1985).

\bibitem{Haldane85}  F.D.M. Haldane, and E.H. Rezayi, Phys. Rev. B {\bf 31}, R2529 (1985).

\bibitem{rrhaldane}  E.H.  Rezayi, and F.D.M. Haldane,  Phys. Rev. Lett. {\bf 84}, 4685  (2000).

\bibitem{Westerberg} E. Westerberg, and T.H. Hansson, Phys. Rev. B {\bf 47}, 16544 (1993).

\bibitem{chui86} S.T. Chui, Phys. Rev. Lett. {\bf 56}, 2395  (1986).

\bibitem{Schrieffer} W.P. Su, and J.R. Schrieffer, Phys. Rev. Lett. {\bf 46}, 738 (1981).

\bibitem{Pan03} W. Pan, H.L. Stormer, D.C. Tsui, L.N. Pfeiffer, K.W. Baldwin, and K.W. West,  Phys. Rev. Lett. {\bf 90}, 016801 (2003).

\bibitem{hierarchy} F.D.M. Haldane, Phys. Rev. Lett. {\bf 51}, 605 (1983); B.I. Halperin, Phys. Rev. Lett. {\bf 52}, 1583, 2390(E) (1984).

}
\end{thebibliography}
\end{document}